\documentstyle[12pt]{article}
\begin{document} 

\title{Relational Relativity}
\author{Amir H. Abbassi$^1$$\;\;\;\&\;$ Amir M. Abbassi$^2$ \\
\textit{\small $^1$Department of Physics, School of Sciences,
Tarbiat Modarres University, }\\
\textit{\small P.O.Box 14155-4838, Tehran, Iran.}\\
\textit{\small E-mail: ahabbasi@modares.ac.ir}\\
$\;$\\
\textit{\small $^2$ Department of Physics, Faculty of Sciences,
Tehran University,}\\
\textit{\small North Kargar Ave., Tehran  14394, Iran.}\\
\textit{\small E-mail: amabasi@khayam.ut.ac.ir}}
\date{February 2002}
\maketitle

\begin{abstract}
According to a simple model of inertia a Machianized theory of
special and general relativity named as relational relativity is presented.  

\bigskip
\noindent Keywords: Inertia, Mach's Principle, Relativity,  

\end{abstract}

\begin{center}
\bf{I.$\;$ Introduction}
\end{center}

The famous issues of Newtonian absolute space and time were followed
by many constructive critiques of relationalists. The most efficient works
of this kind were due to the Ernst Mach, the contemporary physicist and 
philosopher[1]. Various aspects of Mach's ideas concerning the motion
, from Newton's bucket to quantum gravity,
have been collected in the proceeding of the conference
held at T\"ubingen (July 1993) for this purpose[2].
Also among recent references the reader is refered to the
works of Assis and Ghosh [3,4].\\
In his critique of Newtonian mechanics(NM), 
Mach arrived at the following two conclusions:
\begin{itemize}
\item[i)]- Only the relative motion of a body with respect to other bodies 
is observable, not motion with regard to absolute space.
\item[ii)]- The inertial motion of a body is influenced by all the masses 
in the Universe.
\end{itemize}
  
To appreciate fully these two physically pleasant ideas we have applied 
them in a proposed classical model of inertia[5]. 
In this model we consider  the inertia as a real two body interaction.
For a system of two particles 1 and 2, in an arbitrary non-rotating
reference frame $S$, this force  
is proportional to the difference of their accelerations with
respect to $S$ and to the inertial charges of each particle as follows:
\begin{equation}
\vec {F}_{inertia}=\mu c_1 c_2(\vec a_1 - \vec a_2)
\end{equation}\label{1}
Here $\mu$ is a coupling constant , $c_1$ and $c_2$ 
represent the inertial charges of particles 1 and 2 respectively and $(\vec a_1$ and $\vec a_2)$ are their 
accelerations with respect to $S$. In a system consisting of N particles the total 
force imposed on the $i$th particle is :  
\begin{equation}
\vec F_i =\mu c_i \sum\limits_{j=1}^{N} c_j (\vec a_i -
\vec a_j )  \label{2}
\end{equation} 
where the index $i$ referes to the particle $i$ and the summation
is over all particles. By defenition in the real world the inertial charge and the Newtonian
inertial mass of a particle are related as follows:
\begin{equation}
m_i = \mu c_i  \sum\limits_{j=1}^{all} c_j  .\label{3}
\end{equation} 
Summation is taken over all particles in the Universe. This means that the inertial mass of a 
particle (say labelled j) depends on its own feature ($c_j$) and a global 
effect of all particles in the world($\sum\limits_{j=1}^{all}c_j$) . This may be considered as 
a simple formulation of Mach's idea concerning inertia.
Since local inhomogenities have no observed effect on the inertial
mass then it is accepted that the inertial mass is determined 
by the global structure of the Universe and this is exactly
expressed by Eq.(3). Rewriting Eq.(2)
in terms of the inertial masses yields a modified form of the Newton's second law, i.e.,
\begin{equation}
\vec F_i = m_i \left ( \vec a_i - \frac{\sum\limits_{j=1}^{all}m_j \vec a_j}{\sum\limits_
{j=1}^{all} m_j} \right )    \label{4}
\end{equation}  
As it is evident these equations are invariant under a more general transformation than Galilean's.
These transformations  may be called generalized Galilean transformations
with the form:
\begin{equation}
\left \{
\begin{array}{l}
t^{^\prime}=t\\
  \vec a^{^\prime}=\vec a -\vec b\\
\vec u^{^\prime}=\vec u-\vec b t-\vec v\\
\vec x^{^\prime}=\vec x -\frac 12\vec b t^2-t\vec v +\vec x_0
\end{array}
\right .\label{5}
\end{equation} 
where $\vec b$, $\vec v$ and $\vec x_0$ are constant acceleration ,velocity
and position of $S^{^\prime}$ with respect to $S$ at $t=0$ respectively. 

Eq.(4) satisfies full Machian aspects.
The one which is of  interest is that  the so-called absolute space is just the frame attached to the 
center of mass of the Universe in which the Newtonian second law,
$\vec F_i = m_i \vec a_i$ is recovered. 
The main feature of this model from a Machian point of view is its relational nature, 
so that the presence of each particle in the Universe and its location relative to the 
others determine the inertial reference frames.
It is seen that Eqs. (2) and (4) also satisfy Newton's third law automatically.
For a two particle system we have $\vec F_1=-\vec F_2$ and for a system with N particles
they make $\sum\limits_{i=1}^{N} \vec F_i =0$.

We may also extend this model to  gravity. Equivalence principle 
here means that the source of inertia and gravitation is the same. Let us define
the gravitational force between two particles of inertial charges $c_1$ and $c_2$ as:
\begin{equation}
|\vec F_G|=\frac{\mu^2 c_1 c_2}{|\vec r_{12}|^2}  \label{6}
\end{equation}
Then we can express gravitational constant $G$ in terms of inertial charges $c_i$s;
\begin{equation}
G=\left(\sum\limits_{j=1}^{all} c_j\right)^{-2}  \label{7}
\end{equation}
or 
\begin{equation}
G=\frac{\mu}{\sum\limits_{j=1}^{all} m_j}  \label{8}
\end{equation}\\
Eqs. (7) and (8) show that $G$ as a global effect is resulted from all inertial charges
and we may infer that $\sum\limits_{j=1}^{all} c_j$ is finite. 
 According to the Mach's ideas the so-called
physical constants (including G) should be determined from global features
of the Universe.
Thus Eqs. (7) and (8) reveal the very feature of a good Machian model. \\

The Lagrangian function
from which Eq.(4) may be extracted is simply obtained by following the canonical procedure of D'Alembert's principle.
Starting from Eq.(4) and restricting ourselves to systems for
which the virtual work of the forces of constraint vanishes
we obtain 
\begin{equation}
\sum\limits_i \left [\vec F_i- m_i \left ( \vec a_i - \frac{\sum\limits_{j=1}^{all}m_j \vec a_j}{\sum\limits_
{j=1}^{all} m_j} \right )\right ]\cdot \delta \vec r_i =0 ,
\end{equation}\label{9} 
which is the new form of D'Alembert principle. Here $\delta \vec {r_i}$s are infinitesimal changes of coordinates
as the result of virtual displacement of the system.
This leads to the result just
like the one in ordinary NM except that the kinetic energy $T$ that is equal to
$\sum\limits_{i} \frac 12 m_i {v_i}^2$ should be replaced by:  
\begin{eqnarray}
T&=& \sum\limits_{i} \frac 12 m_i v_i^2 - \frac{\left [ \sum\limits_{i}
 m_i v_i\right ] ^2}{2\sum\limits_i m_i} \\
&=&\frac 14 \sum\limits_{i} \sum\limits_{j} m_i m_j \frac{(\vec v_i -\vec
v_j)^2}{\sum\limits_{k} m_k} \nonumber \label{10}
\end{eqnarray}
Indeed the difference is the second term in the first row and is just the kinetic energy of the center of mass which is canceld out in this model.  This is well justified when is applied in cosmology. Where we are dealing with the whole Universe, motion and kinetic energy 
of its center of mass have no physical meaning.\\

The new form of $T$ as a function of the magnitude of relative 
velocities of particles has a scalar invariant manner. 
Then as an other advantage in this model the Lagrangian and Hamitonian
of a system  are scalar invariants from point of view of a 
nonrotating observer.
\begin{equation}
L=\frac 14 \sum\limits_i\sum\limits_j m_im_j
\frac{(\vec v_i - \vec v_j)^2}{\sum\limits_k m_k}-V(r_{ij})  \label{11}
\end{equation}
It is noticeable that in a different way to obtain a relational
NM Eq.(11) has been proposed by Lynden-Bell[6,7].\\

We may summarize the Machian features of this model as follows:
\begin{enumerate}
\item The relational nature of this model is so that by considerng 
relational distances there is no need to assume absolute space or 
inertial frame. Indeed the so-called inertial frame is the
frame attached to the center of mass of the Universe. Then existance 
of each particle and its location with respect to others determine 
the inertial frames.
\item Inertial mass of each particle depends on its own inertial charge
and the sum of inertial charges of all particles in the world.
It is not a natural constant , and may change whenever the total inertial charge 
of the world undergoes any change(e.g. in pair production era).
\item Gravitational constant $G$ is related to the sum of all
inertial charges existing in the Universe and as a global effect
each individual paticle shares in its construction. Just like
inertial mass , this may be changed whenever the total inertial charge 
of the world faces with changes.
\item The concept of energy in this model is independent of
measuring reference frame and is an invariant scalar quantity.
\item For an empty universe it does predict no structure.   
\end{enumerate}
 
Collection of these features in the above model 
provides us a suitable guide to continue and achieve a modified theory of relativity, i.e. a theory of relativity 
without any non-Machian shortcoming, what we may call as relational relativity(RR). As a first step toward RR it is convinient to begin with special relativity (SR).
 
\begin{center}
{\bf II.$\;$ Relational Special Relativity}
\end{center}

At the begining it should be noticed that
according to the Eqs. (7) and (8) it is possible to assume 
a world without inertia via vanishing the coupling constant $\mu$,
but the assumption of a world without gravitation is physically impossible.
Then the subject of special relativity because of its ignorance of
gravitation is under question and cannot be considered as a global theory from a Machian standpoint. In spite of this we try to present a 
relational special theory of relativity.\\
 
Although Michelson-Morley experiment rejects the concept of ether but
SR still is based on the same assumption of the existence of absolute 
space and preference of inertial frames as NM. In a relational
approach we may remove the need for absolute space in SR. To do this
task some preliminary remarks should be mentioned.\\
In NM the Lagrangian of a free particle is just the kinetic energy
and its action is 
\begin{equation}
S=\int{dt (\frac12 m\dot{x}^2)}.     
\end{equation}  \label{12}
where $\dot x$ is the velocity of the particle with mass $m$.
In SR this is changed to the following form
\begin{equation}
S=-m\int{dt(1-{\dot{x}}^2)^{\frac12}} =-m\int{ds}     
\end{equation}\label{13}
so that in low velocity limit $(\dot x\ll 1)$  
the equation of motion returns to the Newtonian form.
Other form of this relation in terms of  space-time metric is
\begin{equation}
S=-m\int{dt\left (g_{\mu\nu} \frac{dx^{\mu}}{dt}\frac{ dx^{\nu}}{dt}\right )^{\frac12}}.   
\end{equation}\label{14}
 That is the Lagrangian is as follows
\begin{equation}
L=-m(g_{\mu\nu}\dot{x}^\mu \dot{x}^\nu)^{\frac12}     
\end{equation}\label{15}
where $g_{\mu\nu} = \eta_{\mu\nu}$ i.e., just the Minkowski metric.\\
According to the definition of canonical momentum we have
\begin{equation}
p_\alpha = \frac{\partial L}{\partial {\dot{x}}^\alpha} =
-\frac{m \eta_{\alpha\mu} {\dot x}^\mu}{(\eta_{\mu\nu}
{\dot x}^\mu {\dot x}^\nu)^\frac12} = -m \eta_{\alpha\mu} u^\mu   
\end{equation} \label{16}
where by definition $\frac{dx^\mu}{ds}=u^\mu$.\\
Then the equation of motion has the form
\begin{equation}
m\dot{u}_\alpha =0.      
\end{equation} \label{17}

For a system of N particles with masses $m_a \,\, , a=1,2,..N$
is made by defining the action as
\begin{equation}
S=-\sum\limits_{a=1}^{N}  m_a \int(\eta_{\mu\nu}\frac{dx^\mu_a}{d p}\frac{\partial
 x^\nu_a}{dp})^\frac12 dp      
\end{equation} \label{18}
where $p$ is an affine parameter and the Lagrangian is: 
\begin{equation}
L=-\sum\limits_{a=1}^{N} m_a (\eta_{\mu\nu}\frac{dx^\mu_a}
{dp}\frac{dx^\nu_a}{dp}
)^\frac12 .          
\end{equation}\label{19}

We should add two other primary remarks about
geometrical and physical points. With physical point we mean a point mass
but a geometrical point need not contain any matter. We should 
insist in this fact that a distance measurement is only made between
two physical points. So in presenting the line element definition instead of
measuring the distance of physical points with respect to an arbitrary
origin we should define it in terms of the distance between physical points
(or physically significant points e.g. center of mass of a system).
Certainly this definition has higher Machian(Relational) validity. Now from this point of view let us define the line element $ds^2_a$
for a noninteracting N particle system as: 
\begin{equation}
ds^2_a=\eta_{\mu\nu}\left ( d{x_a}^\mu - \frac{\sum\limits_b m_b d{x_b}^\mu}
{\sum\limits_b m_b}\right )\left ( d{x_a}^\nu - \frac{\sum\limits_b m_b d{x_b}^\nu}
{\sum\limits_b m_b}\right )       
\end{equation}\label{20}
where index ($a$) refers to the particle labeled ($a$).
Then the related action and Lagrangian are:
\begin{equation}
S=-\sum\limits_a m_a \int [ \eta_{\mu\nu}(\frac{dx^\mu_a}{dp} -\frac{
\sum\limits_b m_b \frac{dx^\mu_b}{dp}}{\sum\limits_b m_b})(\frac{dx^\nu_a}{dp}
 -\frac{\sum\limits_b m_b \frac{dx^\nu_b}{dp}}{\sum\limits_b m_b}) ]
^\frac12 dt   
\end{equation}\label{21}
\begin{equation}
L=-\sum\limits_a m_a \left [\eta_{\mu\nu}\left (\frac{dx^\mu_a}{dp} -\frac{dx^\mu_{cm}}{dp}\right )
\left (\frac{dx^\nu_a}{dp}-\frac{dx^\nu_{cm}}{dp}\right )\right ]^\frac12 . 
\end{equation}\label{22}
The canonical momentum of the kth particle is
\begin{eqnarray}
(p_k)_\alpha =\frac{\partial L}{\partial \frac{dx^\alpha_k}{dp}}&=&
-\eta_{\alpha\nu} m_k \frac{(\frac{dx^\nu_k}{dp}
-\frac{dx^\nu_{cm}}{dp})}{\frac{ds_k}{dp}} \nonumber \\
&=&-\eta_{\alpha\nu} m_k (u_k^\nu - u_{cm}^\nu)
\nonumber \\
&=& -m_k ((u_k)_\alpha - (u_{cm})_\alpha) .  
\end{eqnarray}\label{23}
Since $\frac{\partial L}{\partial x_k^\alpha}=0$ and
$\frac d{dp}(p_k)_\alpha = -m_k \eta_{\alpha\nu}
(\frac{du^\nu_k}{dp} -\frac{du^\nu_{cm}}{dp})$ , 
then the equation of motion of the kth particle is
\begin{equation}
\vec{\frac{du_k}{dp}}-\vec{\frac{du_{cm}}{dp}}=0 
\end{equation} \label{24}
which is just the same as the modified form (4) in the Newtonian limit. 

The Lagrangian (22) is written without any coordination 
with respect to a priori fixed virtual absolute space and these
are particles by their own relative locations that determine
it. This is free from that non-Machian aspects 
suffering the standard SR. So we may call the relativistic theory
based on this Lagrangian as relational special relativity.

\begin{center}
{\bf III.$\;$ Relational General Relativity}
\end{center}

It seems the same approach may be followed to obtain the 
relational GR. But this is not so straightforward. Because to 
extrapolate this result to GR,
i.e. to change the Minkowskian flat space-time
($\eta_{\mu\nu}$) into the Riemannian curved space-time
($g_{\mu\nu}$), care should be taken of dealing with
vector quantities. Summation of the vectors in this case
needs parallel transportation of them
which in turn requires to define the path of transportation 
for each of them. To achieve a relational theory
of GR it requires choosing another strategy with some
different approach as follows.

Initially we remark the center of mass(CM) concept in NM.
With the help of this concept in the Eucledian space NM 
of a single particle can be extrapolated and be applied to
a system with N particles. The classical meaning of CM 
losses its uniqueness when enters in the realm of
relativity so that different observers find different points
as CM of a given system.  The important point
worthy to notice about CM is its dual character
from a Machian point of view so that 
despite of its great value 
as a technical tool to present the relational motion
on the other hand as a point in which total mass of 
the system is located and its motion is to be considered
is quite anti-Machian concept. A single point
has no motion and no inertia. 

Turning back to the NM we may define the center of inertial charge
(CI). By definition:
\begin{eqnarray}
X^\mu_{CI}&=&\frac{\sum\limits_i\sum\limits_j c_i c_j (x^\mu_i+x^\mu_j)}
{2(\sum\limits_j c_j)^2} \nonumber \\ 
&=&\frac{\sum\limits_i m_ix^\mu_i}{\sum\limits_i m_i}=X^\mu_{CM}
\end{eqnarray}\label{25}
where $X^\mu_{CM}$ and $X^\mu_{CI}$ are coordinates of CM and CI 
respectively. As it is evident the concept of CI has also a 
mutually relational content between particles.\\
Now it is easy to show that the result (4) may be obtained
with the help of Lagrangian formalism in NM and imposing
the following condition on CI;
\begin{equation}
\delta X^\mu_{CI}\equiv 0
\end{equation}\label{26}
Because 
\begin{equation}
\delta X^\mu_{CI}=\sum\limits_i m_i\delta x^\mu_i =0 \;,
\end{equation}\label{27}
and imposing this by using the method of undetermined 
Lagrangian multipliers in variations of the action of 
a system with N noninteractiong particles , yields:
\begin{equation}
\delta I=\sum\limits_n\int dt[m_n\ddot x^\mu_n+fm_n]\delta x^\mu_n\equiv 0
\end{equation}\label{28}
where the coefficient $f$ is determined as follows:
\begin{equation}
f= -\frac{\sum\limits_n m_n \ddot x^\mu_n }{\sum\limits_n m_n}
\end{equation}\label{29}
Thus the equation of motion (4) is obtained. 
Also with consideration of the condition (27) in variation of
the action (18)in special relativity the result (22) may be obtained.

To remove the Machian objection to the concept of CM 
the following condition as a Machian condition may be imposed to the variations of the dynamical variables $x^\mu_\nu$ of the system.
Let us first define $X^\mu$ as
\begin{equation}
X^\mu\equiv\sum\limits_n m_n x^\mu_n .
\end{equation}\label{30}
Of course $X^\mu$ is not a vector quantity and depends to the 
chosen reference frame. Let us denote its variations with$\delta X^\mu$:
\begin{equation}
\delta X^\mu \equiv \sum\limits_n m_n \delta x^\mu_n
\end{equation}\label{31}
Here$\delta X^\mu$ is not vector while $\delta x^\mu_n$s are vectors.
Similarly $\delta X_\mu$ is defined as follows:
\begin{equation}
\delta X_\mu \equiv \sum\limits_n m_n g_{\mu\lambda}(x_n) \delta x^\lambda_n
\end{equation}\label{32}
 
Now as a Machian principle we postulate that allways $\delta X_\mu$
vanishes(lower index is chosen only for convenience).
This means that variations of dynamical variables $x^\mu_n$ are 
under the following condition:
\begin{equation}
\sum\limits_n m_n g_{\mu\lambda}(x_n)\delta x^\lambda_n =0
\end{equation}\label{33}

Despite of this fact that (33) is not a covariant condition
we can make the best use of it to find at least a clue for the
geodesic equations in GR.   \\

Matter action for a system consisting of $N$ particles with
masses $m_n$ is given by the following form in GR;
\begin{equation}
I=\sum\limits_n m_n \int dp\left(g_{\mu\nu}(x_n(p))
\frac{dx^\mu_n (p)}{dp} \frac{dx^\nu_n}{dp}\right )^{\frac 12}
\end{equation}\label{34}
where $p$ is some quantity that simultaneously parametrizes all
the space-time trajectories of the various particles.

Variation of the action (34) due to an infinitesimal variation
in the dynamical variables $x^\mu\rightarrow x^\mu(p)+
\delta x^\mu(p)$ is given by:
\begin{eqnarray}
\delta I=\frac 12 \sum\limits_n &m_n&\int dp[g_{\mu\nu}
(x_n (p))\frac{dx^\mu_n (p)}{dp} \frac{dx^\nu_n (p)}{dp}]^{-\frac 12} \nonumber \\
&\;&\times\left\{ 2g_{\mu\nu}(x_n(p))\frac{dx^\mu_n (p)}{dp}
\frac{d\delta x^\nu_n (p)}{dp} \right.\nonumber \\
&\;&+\left. \left ( \frac{\partial g_{\mu\nu}(x)}{\partial x^\lambda}\right )_{x=x_n(p)} \frac{dx^\mu_n (p)}{dp}\frac{dx^\nu_n}{dp}\delta
x^\lambda_n (p)\right \}
\end{eqnarray}  \label{35}
It is convenient to change variables of integration (35) from $p$
to the $\tau_n$ (the proper time of the particle $n$) defined by:
\begin{equation}
d\tau_n\equiv (g_{\mu\nu} dx^\mu_n dx^\nu_n)^{\frac 12}
\end{equation}\label{36}
So the integral in (35) may be written in a simpler form:
\begin{eqnarray}
\delta I=\frac 12 \sum\limits_n m_n \int d\tau_n\left \{
2g_{\mu\lambda}(x_n)\frac{dx^\mu_n}{d\tau_n}\frac{d\delta x^\lambda
_n}{d\tau_n}\right. \nonumber \\
+\left. \frac{\partial g_{\mu\nu}(x_n)}{\partial x^\lambda_n}
\frac{dx^\mu_n}{d\tau_n}\frac{dx^\nu_n}{d\tau_n}\delta x^\lambda_n
\right \}
\end{eqnarray}\label{37}
 
Finally integration by parts of the first term in (37) with the condition that 
$\delta x^\mu(\tau_n)$ vanishes on the boundaries of integration 
yields that :
\begin{equation}
\delta I= \sum\limits_n \int d\tau_n g_{\mu\lambda}(x_n)
\left\{ m_n (\frac{d^2x^\mu_n}{d\tau_n^2}+\Gamma^\mu_{\rho\sigma}
\frac{dx^\rho_n}{d\tau_n}\frac{dx^\sigma_n}{d\tau_n})\right\}
\delta x^\lambda_n
\end{equation}\label{38}
where $\Gamma^\mu_{\rho\sigma}$ are the second type Christoffel symbols. 
Then according to the principle of stationary action, 
$\delta I$ vanishes for general variations 
in the dynamical variables $\delta x^\lambda_n$ if and only if the 
dynamical variables obey the geodesic equations:
\begin{equation}
\frac{d^2x^\mu_n}{d\tau_n^2}+\Gamma^\mu_{\rho\sigma}
\frac{dx^\rho_n}{d\tau_n}\frac{dx^\sigma_n}{d\tau_n}=0
\end{equation}\label{39}

Now we repeat the above standard process with consideration
of the Machian condition (33) to achieve the equations of motion.
To impose the mentioned condition with the method of undetermined
Lagrangian multipliers it is enough only to add the following term
to the variations of the action (34)
\begin{equation}
\int dp f^\mu\sum\limits_n m_n g_{\mu\lambda}(x_n)\delta x^\lambda_n
\end{equation}\label{40}
where $f^\mu$s are undetermined coefficients and just as in (34)
parameter $p$ is an arbitrary quantity which simultaneously
parametrizes the space-time trajectories of different particles.
Then we have:
\begin{equation}
\delta I=\sum\limits_n\int dp\left\{ m_n g_{\mu\lambda}(x_n)
[\frac{\partial p}{\partial \tau_n}(\frac{d^2x^\mu_n}{dp^2}
+\Gamma^\mu_{\rho\sigma}\frac{dx^\rho_n}{dp}\frac{dx^\sigma_n}
{dp})+f^\mu]\right\}\delta x^\lambda_n =0
\end{equation}\label{41}
With $f^\mu$s determined as:
\begin{equation}
f^\mu=-\frac{\sum\limits_n m_n\frac{\partial p}{\partial\tau_n}
(\frac{d^2x^\mu_n}{dp}\frac{dx^\sigma_n}{dp})}{\sum\limits_n m_n}
\end{equation}\label{42}
Because of the mean operation over all particles $f^\mu$ 
is a global quantity.\\

Therefore by inserting the value of $f^\mu$ the Machianized form
or the relational form of the geodesic equations of motion are derived
as follows:
\begin{equation}
\frac{d^2x^\mu_n}{dp^2}-\frac{\sum\limits_j m_j\frac{\partial \tau_n}
{\partial \tau_j}\frac{d^2x^\mu_j}{dp^2}}{\sum\limits_j m_j}
+\Gamma^\mu_{\rho\sigma}\frac{dx^\rho_n}{dp}\frac{dx^\sigma_n}{dp}
-\frac{\sum\limits_j m_j\frac{\partial\tau_n}{\partial\tau_j}
\Gamma^\mu_{\rho\sigma}\frac{dx^\rho_j}{dp}\frac{dx^\sigma_j}{dp}}
{\sum\limits_j m_j}=0
\end{equation}\label{43}
    
It reveals that in the weak field limit the equations
(43) correspond with the Newtonian one , because the Christoffel symbols
vanish and parameters $\tau_n$ in this limit are all the same and are 
equal to $t$, then Eq.(43) reduces to the modified Newtonian
form Eq.(4).

Now according to the relational result (43) we may propose 
the covariant form of the geodesic equations as follows:
\begin{equation}
\frac{d^2x^\mu_n}{dp^2}+\Gamma^\mu_{\rho\sigma}
\frac{dx^\rho_n}{dp}\frac{dx^\sigma}{dp}-
\frac{\sum\limits_j m_j\frac{\partial\tau_n}{\partial\tau_j}U^{x_n}_{x_j}
(\frac{d^2x^\mu_j}{dp^2}+\Gamma^\mu_{alpha\beta}\frac{dx^\alpha_j}
{dp}\frac{dx^\beta_j}{dp})}{\sum\limits_j m_j}=0
\end{equation}\label{44}
where $U^{x_n}_{x_j}$is the parallel transportation operator
from the location of the jth particle to the location of the nth one. \\

\begin{center}
{\bf IV.$\;$ Remarks}
\end{center}

We are now staying at a standpoint that may return to the famous question 
that ``whether the formalism of general relativity and the 
Einstein equations are perfectly Machian?'' and have a strictly 
positive answer to it. Checking the Machian(or anti-Machian)
aspects of GR we notice that;
\begin{enumerate}
\item By now
in front of the basic question that why the Einstein field
equations have nontrivial solution flat space $R_{\mu\nu}=0$
for empty universe we had to resort to the boudary condition 
reasons. Hereafter, with what we have find about inertia it is
seen that the Einstein field equations $\frac{c^4}{8\pi G}R_{\mu\nu}=
(T_{\mu\nu}-\frac 12 g_{\mu\nu}T)$ predict $0=0$ (instead of 
$R_{\mu\nu}=0$)for empty universe. For assuming vacuum $T\;,\;
T_{\mu\nu}=0$ makes the RHS of the field equations 
to be equal zero and on the other side the coupling constant
appears on the LHS as $G^{-1}$, which in turn according to the
relations (7) and (8) depends on the existence of all particles
in the universe, $G\propto\frac 1{\sum\limits_i m_i}$, so
for the empty universe $\sum\limits_i m_i=0$ and thus the
field equations yield to $0=0$, that is a perfectly Machian result. 
\item For a world with a single particle, although the 
field equations based on the presented model of inertia 
predict a solution that is independent of inertial charge
and merely depending to the coupling constant $\mu$.
But for its geodesic equation the relations (43) and (44)
yield to the result $0=0$, that means denying any motion 
for a single particle , an ideal result from a Machian
point of view.
\end{enumerate} 
\newpage
\begin{center}
{\bf Acknowledgements}
\end{center}

A.M.A. would like to thank Prof. J. Barbour
for his comments and encouragements. 


\bigskip 

 

\begin{thebibliography}{1}
\bibitem{1} Mach, E., {\it The science of Mechanics}, 
(The Open Court Publishing Co., 1974).
\bibitem{2} Barbour,J., Pfister,H. (Eds.)(1995).{\it Mach's principle
: From Newton's Bucket to Quantum Gravity}, Birkhauser, Boston.
\bibitem{3} Andre K.T. Assis, Relational Mechanics, Apeiron(1999).
\bibitem{4} Amitabha Ghosh, Origin of Inertia: Extended Mach's
principle and cosmological consequences, Apeiron(2000).
\bibitem{5} Abbassi, A. H., Abbassi, A. M., A Modified Theory
of Newtonian Mechaincs, J.Sci.I.R.Iran,Vol.7,No.4,277-279,1996.
(arXiv:physics/0006021).

\bibitem{6} Lynden-Bell, D., A Relative Newtonian Mechanics,
ref[3], pp172-178.

\bibitem{7} Lynden-Bell,D.,Katz,J., Classical mechanics without absolute space, PRD, Vol.52,No.12,7322-7323,1995.


\end{thebibliography}
\end{document}